\documentstyle[12pt]{article} %available in the standard LaTex package.
\normalbaselines
\def\tr{\mathop{\rm tr}\nolimits}
\def\id{\mathop{\rm Id}\nolimits}
\begin{document}
\bibliographystyle{unsrt}
\begin{center}
{\large \bf ASYMPTOTIC QUANTUM PARAMETER ESTIMATION
\\[2mm]
IN SPIN 1/2 SYSTEM}\\[7mm]
Masahito Hayashi \\
{\it Department of Mathematics, Kyoto University, Kyoto 606-01, Japan \\ 
e-mail address: masahito@kusm.kyoto-u.ac.jp }\\[5mm]
\end{center}
\begin{abstract}
For the precise estimation of the unknown quantum state, 
the independent samples should be prepared.
Can we reduce the error of the estimation by the measurement using
the quantum correlation between every sample?
In this paper, this question is treated in the parameter estimation for the unknown state.
\end{abstract}
\section{Introduction}
There are more and more papers about quantum state estimation and state reconstruction and tomography.
But, in these paper there are few descriptions about the minimization of the 
estimation error and a asymptotic evaluation of it \cite{Jon,BAD,Da}.\par

When we estimate the unknown state precisely,
we must prepare many independent identical samples of the unknown state \cite{DY}.
For the estimation of the unknown state,
if we use the quantum correlation between every sample,
can we reduce the error of the estimation?
Massar and Popescu answered this question in the case of pure states on 
spin 1/2 system about \underline{\it finite samples} 
in the sense of Bayes inference \cite{smsp}.
On the other hand, Hayashi answered this question in the case of pure states on a finite-dimensional quantum system about the 
\underline{\it asymptotic efficiency}
in the sense of Bayes and minimax \cite{Haya3}.
To Massar and Popescu's result, in the case of pure states
on spin 1/2 system, the optimal measurement for finite samples must necessarily view these samples as a single composite system i.e. the optimal measurement cannot be realized by separate measurements on each system.
But, to Hayashi's result, the minimum error can be asymptotically attained by separate measurements on each system in Bayes and minimax in the first order.

In this paper, we treat this problem under the quantum parameter estimation,
in which we assume that the unknown state is included in a state family 
characterized by finite parameters.
Therefore we compare the minimum error between the two case 
under the local unbiasedness conditions.
In the estimation for a coherent light in thermal noise, the minimum
error in the case of these samples viewed as a single composite system
can be attained by separate measurements on each system.
But in spin 1/2 system full family, it cannot be attained.

\section{Asymptotic Quantum Parameter Estimation}
In this paper, we admit that a quantum state is described by a density operator. 
In the quantum parameter estimation, assuming that the unknown state is included
by a certain quantum state family ${\cal S}$ 
parameterized by finite parameters,
we search an efficient estimator to estimate parameters characterizing the unknown state.
\begin{eqnarray*}
{\cal S}:=\{ \rho_{\theta} \in {\cal S}({\cal H}) | \theta = ( \theta^1 , \ldots , \theta^d ) \in \Theta \subset {\bf R}^d \},
\end{eqnarray*}
where the set ${\cal S}({\cal H})$ denotes the set of densities on ${\cal H}$.
The most general description of a quantum measurement probability is given by the mathematical concept of a \underline{\it positive operator valued measure} (POM) on the system state space \cite{Hel,HolP}.
Generally speaking, if $\Omega$ is a measurable space, 
a measurement $M$ satisfies the following:
\begin{eqnarray*}
M(B)=M(B)^* \ge 0, M( \emptyset ) = 0, M(\Omega) = \id \hbox{ \rm on } {\cal H},
\hbox{ \rm for any Borel } B \subset \Omega. \\
M( \cup_i B_i ) = \sum_i M( B_i), \hbox{ \rm for } B_i \cap B_j = \emptyset
(i \neq j), \{ B_i \} \hbox{ \rm is a countable subsets of } \Omega.
\end{eqnarray*}

In the quantum parameter estimation, we often treat the local unbiasedness conditions
characterizing locally efficient estimators.
A measurement $M$ is called \underline{\it locally unbiased} at $\rho_{\theta_0}$ if
\begin{eqnarray*}
\int_{{\bf R}^d} \hat{\theta}^i \tr M ( \,d \hat{\theta} ) \frac{ \partial \rho_{\theta}}{\partial \theta^j} \Bigl |_{\theta = \theta_0 } = \delta^i_j 
,~
\int_{{\bf R}^d} \hat{\theta}^i \tr M ( \,d \hat{\theta} ) \rho_{\theta_0 } = 
\theta_0^i .
\end{eqnarray*}
There are two quantum analogues of the logarithmic derivative.
One is the symmetric logarithmic derivative (SLD) $L_i$ defined as satisfying
$\frac{1}{2}(L_i \rho + \rho L_i) = \frac{ \partial \rho_{\theta}}{\partial \theta^i} \Bigl |_{\theta = \theta_0 }$.
The another is the right logarithmic derivative (RLD) $\tilde{L_i}$ defined as
$\tilde{L_i}:=\frac{ \partial \rho_{\theta}}{\partial \theta^i} \Bigl |_{\theta = \theta_0 }\rho^{-1}$.
Therefore, there exist two quantum analogues of the Fisher information matrix.
One is the SLD Fisher information matrix $J_{\rho}$ defined as
$J_{\rho}:= \tr [ \frac{1}{2}(L_i L_j + L_j L_i) \rho]_{i,j}$.
The another is the RLD Fisher information matrix $\tilde{J}_{\rho}$ defined as
$\tilde{J}_{\rho}:= [\tr \tilde{L}_i^* \tilde{L}_j\rho]_{i,j}$.
Thus, there are two quantum analogues of the Cram\'{e}r-Rao inequality as follows.
If a measurement $M$ is locally unbiased at $\rho=\rho_{\theta_0}$,
then we obtain the following inequalities:
\begin{eqnarray}
V_{\rho}(M) &\ge& J_{\rho}^{-1} \hbox{ (the SLD Cram\'{e}r-Rao Inequality)} \label{SLDi} \\
V_{\rho}(M) &\ge& \tilde{J}_{\rho}^{-1} \hbox{ (the RLD Cram\'{e}r-Rao Inequality)} , \label{RLDi}
\end{eqnarray}
where $V_{\rho}(M)$ denotes the covariance matrix 
$[\int_{{\bf R}^d} (\hat{\theta}^i -\theta_0^i)(\hat{\theta}^j-\theta_0^j) \tr M(\,d \hat{\theta}) \rho ]_{i,j}$
of a measurement $M$ at a state $\rho$.
If a state $\rho$ is faithful (nondegenerate) and SLDs $L_1, \ldots , L_d$
are noncommutative with each other,
then the lower bound of the SLD Cram\'{e}r-Rao Inequality (\ref{SLDi}) cannot be 
attained.
Therefore, we search the estimator which minimizes $\tr G V_{\rho}(M) $ under
the local unbiasedness conditions, where $G$ is any weight matrix satisfying that
$G_{i,j}=G_{j,i} \in {\bf R}, G \ge 0$.
The minimum value is called the \underline{\it attainable Cram\'{e}r-Rao type bound}
and is denoted by $C(G)$.
From the RLD Cram\'{e}r-Rao Inequality (\ref{RLDi}), we obtain
\begin{eqnarray*}
C(G) \ge \tr \tilde{J}_{\rho}^{-1} G' ,
\end{eqnarray*}
where $G'$ is an Hermite matrix satisfying that $G' \ge 0 , \tr G X = \tr G' X $ for
any real symmetric matrix $X$.
The maximum value $\tr \tilde{J}_{\rho}^{-1} G'$ under the preceding condition
is called \par\noindent the \underline{\it RLD Cram\'{e}r-Rao type bound} and is denoted by $C_R(G)$ 
as the maximum value is the lower bound of the attainable Cram\'{e}r-Rao type bound $C(G)$.

To know the covariance matrixes under the local unbiasedness conditions,
we research the set ${\cal V}_{\rho}$ of covariance matrixes under these conditions.
It is sufficient for
the research of ${\cal V}_{\rho}$ to investigate the set 
$K({\cal V}_{\rho}):= \{ a \in {\cal V}_{\rho} | b \in {\cal V}_{\rho}, b \le a
\Rightarrow b=a \}$
because ${\cal V}_{\rho}= \{ a + b | a \in K({\cal V}_{\rho}) , b \ge 0 \}$.
\par
Next, we assume that we prepare $n$ independent identical samples of the unknown state $\rho_{\theta}$.
The preceding condition is called the quantum independent and identically distribution (i.i.d.) condition.
Under this condition, the quantum state is described by $\rho_{\theta}^{(n)}$ defined by:
\begin{eqnarray*}
\rho_{\theta}^{(n)} :=  \underbrace{\rho_{\theta} \otimes \cdots \otimes \rho_{\theta} }_{n}
\hbox{ on } {\cal H}^{(n)} ,
\end{eqnarray*}
where the composite system ${\cal H}$ is defined as:
\begin{eqnarray*}
{\cal H}^{(n)} := \underbrace{{\cal H} \otimes \cdots \otimes {\cal H}}_{n} .
\end{eqnarray*}
When we use $n$ samples of the unknown state, we may consider the following family ${\cal S}^{(n)}$ called 
the $n$-i.i.d. extended family of ${\cal S}=\{ \rho_{\theta} \in {\cal S}({\cal H}) | \theta \in \Theta \}$:
\begin{eqnarray*}
{\cal S}^{(n)} := \{ \rho_{\theta}^{(n)} \in {\cal S}({\cal H}^{(n)}) | \theta \in \Theta \}.
\end{eqnarray*}
If a measurement $M$ is locally unbiased at $\rho^{(n)}=\rho_{\theta_0}^{(n)}$
in the $n$-i.i.d. extended family,
then we have
\begin{eqnarray}
n V_{\rho^{(n)}}(M) &\ge& J_{\rho}^{-1} \hbox{ (the SLD $n$-i.i.d. Cram\'{e}r-Rao Inequality)} \label{nSLDi} \\
n V_{\rho^{(n)}}(M) &\ge& \tilde{J}_{\rho}^{-1} \hbox{ (the RLD $n$-i.i.d. Cram\'{e}r-Rao Inequality)}  . \label{nRLDi}
\end{eqnarray}
To know the efficiency of using the quantum correlation,
we calculate the infimum value of $n \tr G V_{\rho^{(n)}}(M)$ with respect to 
$(M,n)$
under the condition that
a measurement $M$ is locally unbiased at $\rho^{(n)}=\rho_{\theta_0}^{(n)}$
in the $n$-i.i.d. extended family.
This minimum value is called the \underline{\it asymptotic attainable Cram\'{e}r-Rao type bound} and is denoted by $C_A(G)$.
From the RLD $n$-i.i.d. Cram\'{e}r-Rao Inequality (\ref{nRLDi}) and 
the definitions of $C(G)$ and $C_A(G)$,
we obtain these inequalities:
\begin{eqnarray}
C(G) \ge C_A(G) \ge C_R(G) \label{aar}.
\end{eqnarray}
Let ${\cal V}_{A,\rho}$ be the closure of the set of covariance matrixes of $(M,n)$ 
under the condition that
a measurement $M$ is locally unbiased at $\rho^{(n)}=\rho_{\theta_0}^{(n)}$
in the $n$-i.i.d. extended family, then
it is easily derived that ${\cal V}_{\rho} \subset {\cal V}_{A,\rho}$.

If ${\cal S}$ is a one-parameter family, the attainable Cram\'{e}r-Rao type bound equals the asymptotic attainable Cram\'{e}r-Rao type bound.
If a family ${\cal S}$ consists of pure states,
then the attainable Cram\'{e}r-Rao type bound equals the asymptotic attainable Cram\'{e}r-Rao type bound.
To know the attainable Cram\'{e}r-Rao type bound in pure states, see Matsumoto
\cite{Matsu}.
\section{Examples}
\subsection{quantum thermal states family}
In the estimation for a coherent light in thermal noise,
we estimate the complex parameter $\theta$ of the quantum thermal states family defined as:
\begin{eqnarray*}
{\cal S}:= 
\left\{ \left. \rho_{\theta} := \frac{1}{\pi N}
\int_{{\bf C}} \exp \left( - \frac{| \theta - \alpha | ^2}{N} \right) | \alpha \rangle 
\langle \alpha | \,d^2 \alpha \right| \theta \in {\bf C} \right\}.
\end{eqnarray*}
This family is investigated by Yuen, Lax and Holevo \cite{HolP, YL}.
They calculated the attainable Cram\'{e}r-Rao type bound as follows:
\begin{eqnarray}
C(G)= C_R(G) . \label{saku}
\end{eqnarray}
This bound is attained by a unbiased measurement. 
From (\ref{saku}) and (\ref{aar}), we have
\begin{eqnarray*}
C(G) = C_A(G) = C_R(G) .
\end{eqnarray*}
Therefore, in order to constitute the optimal estimator, 
 we don't have to view these samples as a single composite system in this family.
\subsection{Spin 1/2 system full family}
In spin 1/2 system, to determine the unknown state we estimate
three parameters $(r,\theta,\phi)$ in spin 1/2 system full family defined as:
\begin{eqnarray*}
{\cal S}:= 
\left. \left\{ \rho_{(r,\theta, \phi )} :=
\frac{1}{2} 
\left(
\begin{array}{cc}
1+r \cos \theta & r \sin \theta e^{i \phi} \\
r \sin \theta e^{-i \phi} & 1 - r \cos \theta 
\end{array}
\right )
\right | 
0 \le r \le 1 , 0 \le \theta \le 2 \pi , 0 \le \phi \le 2 \pi 
\right \}.
\end{eqnarray*}
Hayashi derived the following result \cite{Haya1,Haya2}:
\begin{eqnarray}
C(G)&=& \left( \tr \sqrt{ J^{-1/2}_{\rho} G J^{-1/2}_{\rho}} \right)^2 \label{siki1} \\
K({\cal V}_{\rho}) &=& \{ J_{\rho}^{-1/2} W^{-1} J_{\rho}^{-1/2} | \tr W =1 \}.\label{siki2}
\end{eqnarray}
In any 2-parameter normal subfamily of this family, Nagaoka, Fujiwara and Hayashi derived the same equations(\ref{siki1})(\ref{siki2}) \cite{Naga1,FN,Haya1,Haya2}.
In this family the RLD bound can be asymptotically attained i.e. we have
\begin{eqnarray*}
C(G) \,> C_A(G) = C_R(G). 
\end{eqnarray*}
A proof of this claim is too long to show now.
If 
\begin{eqnarray*}
\rho=\rho_{(r,\pi/2,0)}=
\frac{1}{2}
\left(
\begin{array}{cc}
1 & r \\
r & 1 
\end{array}
\right) ,
\end{eqnarray*}
then we have
\begin{eqnarray*}
K({\cal V}_{A,\rho}) = 
\left\{ \left. \left(
\begin{array}{ccc}
1-r^2 & 0 & 0 \\
0 & x+y & z \\
0 & z & x-y 
\end{array}
\right) \right| x= \frac{1}{r^2}\left(1+\sqrt{r^4(y^2+z^2)+r^2} \right) 
\right\} .
\end{eqnarray*} 
\subsection{subfamily $r=r_0$}
If we know the parameter $r=r_0$ of the unknown state in the spin 1/2 system full family,
then we estimate nother two parameters $(\theta,\phi)$ in ${\cal S}_{r=r_0}$ defined as:
\begin{eqnarray*}
{\cal S}_{r=r_0}:= \{ \rho_{(r_0,\theta, \phi )} |  0 \le \theta \le 2 \pi , 0 \le \phi \le 2 \pi  \}.
\end{eqnarray*}
In subfamily $r=r_0$ of spin 1/2 system full family,
we have
\begin{eqnarray*}
C(G) \ge C_A(G) = C_R(G) .
\end{eqnarray*}
If 
\begin{eqnarray*}
\rho=\rho_{(r_0,\pi/2,0)}=
\frac{1}{2}
\left(
\begin{array}{cc}
1 & r_0 \\
r_0 & 1 
\end{array}
\right) ,
\end{eqnarray*}
then 
\begin{eqnarray*}
L_{\theta}= r_0 
\left(
\begin{array}{cc}
-1 & 0 \\
0 & 1 
\end{array}
\right),
L_{\phi}= r_0 
\left(
\begin{array}{cc}
0 & i \\
-i & 0 
\end{array}
\right).
\end{eqnarray*}
Let $G$ as follows:
\begin{eqnarray*}
G=
\left(
\begin{array}{cc}
g_1 + g_2 & g_3 \\
g_3 & g_1 - g_2
\end{array}
\right) \,> 0,
\end{eqnarray*}
then
\begin{eqnarray*}
C(G)=
\frac{2}{r_0^2} \left(g_1 + \sqrt{g_1^2 - g_2^2 - g_3^2} \right) 
\ge
\frac{2}{r_0^2} \left(g_1 + r_0 \sqrt{g_1^2 - g_2^2 - g_3^2} \right) 
= C_A(G) = C_R(G) .
\end{eqnarray*}
The equality establishes iff the state $\rho$ is a pure state i.e. $r_0 =1$.
We have
\begin{eqnarray*}
K( {\cal V}_{\rho})
&=& \left\{ \left. 
\left(
\begin{array}{cc}
x+y & z \\
z & x-y 
\end{array}
\right)
\right| x= \frac{1}{r_0^2}\left(1+\sqrt{r_0^4(y^2+z^2)+1} \right) \right\} = 
\{ J_{\rho}^{-1/2} W^{-1} J_{\rho}^{-1/2} | \tr W =1 \}\\
K ({\cal V}_{A,\rho})
&=&
\left\{ \left. 
\left(
\begin{array}{cc}
x+y & z \\
z & x-y 
\end{array}
\right)
\right| x= \frac{1}{r_0^2}\left(1+\sqrt{r_0^4(y^2+z^2)+r_0^2} \right) \right\} .
\end{eqnarray*}
Therefore, in order to constitute the optimal estimator in this model, 
we have to view these samples as a single composite system except the pure state case.
\subsection{subfamily $\phi=0$}
If we know the parameter $\phi=0$ of the unknown state in the spin 1/2 system full family,
then we estimate the other two parameters $(r,\theta)$ in ${\cal S}_{\phi=0}$ defined as:
\begin{eqnarray*}
{\cal S}_{\phi=0}:= \{ \rho_{(r,\theta, 0 )} | 0 \le r \le 1 ,  0 \le \theta \le 2 \pi \}.
\end{eqnarray*}
In this family,
we have
\begin{eqnarray}
C(G) =  \left( \tr \sqrt{ J^{-1/2}_{\rho} G J^{-1/2}_{\rho}} \right)^2 
\,> \tr J_{\rho}^{-1} G = C_A(G). \label{siki3}
\end{eqnarray}
These equations and this inequality (\ref{siki3}) mean that 
using only finite samples, the bound of the SLD $n$-i.i.d. Cram\'{e}r-Rao Inequality (\ref{nSLDi}) cannot be attained, 
but it can be asymptotically attained.
Therefore, we have $K({\cal V}_{A,\rho})=\{ J_{\rho}^{-1} \}$.
\section{Conclusion}
In spin 1/2 system full parameter family, 
using the quantum correlation reduces 
the covariances under the local unbiasedness conditions.
We need research the same topic in the universal estimation.
\section*{Acknowledgments}
The author wishes to thank to Prof. A. Fujiwara and
Dr. K. Matsumoto for useful discussions on this topic.

\end{document}